\documentclass[12pt ]{article}
\usepackage{jheppub_kim}
\usepackage{amsmath}
\usepackage{amssymb}
\usepackage{dcolumn}
\usepackage{bm}
\usepackage{color}
\usepackage{epsfig}
\usepackage{amsfonts}
\usepackage{graphicx}
\usepackage{subfigure}
   
\begin{document}  
\title{{More Exact Thermodynamic Analysis of Topological Black Holes in $R^2$ Gravity }}

    \author[a,b]{Sudhaker Upadhyay,\footnote{Corresponding author}\footnote{Visiting Associate, Inter-University Centre for Astronomy and Astrophysics (IUCAA) Pune-411007, Maharashtra, India}}
    \author[c]{Jyotish Kumar,}
    \author[d,b]{Dharm Veer Singh,\footnote{Visiting Associate, Inter-University Centre for Astronomy and Astrophysics (IUCAA) Pune-411007, Maharashtra, India}} 
    	 
 \author[e]{Yerlan Myrzakulov,}
  \author[e]{Kairat Myrzakulov,}
 \author[f]{and Abhishek Ashish}

   \affiliation[a]{Department of Physics, K.L.S. College, Nawada, Magadh University, Bodh Gaya, Bihar 805110,  India}
    \affiliation[b]{School of Physics, Damghan University, P.O. Box 3671641167,  Damghan, Iran}
    \affiliation[c]{P.G. Department of Physics,   Magadh University, Bodhgaya, Bihar  824234, India }
        \affiliation[d]{Department of Physics, GLA University, Mathura, Uttar Pradesh 281406, India}
     \affiliation[e]{Department of General \& Theoretical Physics, L. N. Gumilyov Eurasian National University,  Astana, 010008, Kazakhstan}
      \affiliation[f]{Department of Commerce, K.L.S. College, Nawada, Magadh University, Bodh Gaya, Bihar 805110,  India}
\emailAdd{jyotishkumar7137@gmail.com} 
 \emailAdd{sudhakerupadhyay@gmail.com; sudhaker@associates.iucaa.in}
 \emailAdd{veerdsingh@gmail.com} 
 \emailAdd{ymyrzakulov@gmail.com}
  \emailAdd{krmyrzakulov@gmail.com}
\emailAdd{abhishekashish6275@gmail.com}   
\abstract{
This study investigates the thermodynamics of topological black hole solutions in $R^{2}$
  gravity, incorporating the effects of small statistical fluctuations up to first-order corrections. We precisely calculate entropy, internal energy, Helmholtz free energy, specific heat, enthalpy, and Gibbs free energy, accounting for perturbative thermal corrections. Our results reveal that the internal energy of small black holes diverges asymptotically due to these fluctuations. The corrected Gibbs free energy attains asymptotically high values for small horizon radii. In contrast, the equilibrium Gibbs free energy approaches zero.
  Additionally, we assess the stability of the black hole in the presence of these thermal fluctuations. We find that, in contrast to the equilibrium 
  state, the thermal fluctuation introduces a double phase transition to the stability of the black hole.   Our analysis reveals that the influence of fluctuations is notably significant, primarily for small black holes. These findings offer new insights into the thermodynamic properties of topological black holes in the presence of thermal fluctuations.
}
\maketitle
 \section{Introduction}
 Due to renormalisable, scale-invariant and asymptotically
free, quadratic gravity is one of the important gravity models \cite{1,2,3,4,5,6}. Such models generally possess ghosts, but  
the simplest $R^2$ gravity is ghost-free \cite{7}. A scale-invariant quadratic model 
consistent with observations
were proposed as an inflationary model in a much larger physical framework \cite{bh}. This approach involves accounting for the variation of coupling parameters and incorporating an additional scalar field degree of freedom. Other inflationary models align with the most recent observational data \cite{la}. Due to its scale-invariant properties, these theories can be effectively analysed with detectors that do not possess intrinsic scales. For example, a gravitational wave in $R^2$ gravity could be employed as a clock to measure the ratio of its arbitrary frequency relative to another wave.

 Considering black holes as a thermal system does not work without implementing the entropy concept, as the second law of thermodynamics does not hold. 
In this regard, Hawking found that the maximum entropy of black holes has a connection with their event horizon area  \cite{01, II3}. This idea is one of the main pillars of the holographic principles \cite{03,04}. It has been widely advocated that at a specific scale, the maximum entropy of the black holes has corrections, which is attributed to the holographic principle as well \cite{05,06}. 
 The quantum aspects of gravity and small statistical thermal fluctuations around equilibrium are the primary factors contributing to these corrections.
It is widely found that, in the first order, such corrections are logarithmic \cite{II5}, which becomes significant at small black holes.  

Recent research has extensively explored entropy corrections for a variety of black hole types, including charged black holes \cite{16}, rotating black holes \cite{16}, quasitopological black holes \cite{15}, charged massive black holes \cite{17}, $f(R)$ modified black holes \cite{18}, Horava-Lifshitz black holes \cite{19}, and regular black holes \cite{b1, b2, b3, Singh:2022izz,singh}. Despite the significant advancements in understanding these corrections across different black hole models, the impact of small statistical thermal fluctuations on entropy has not yet been examined for topological black holes within the framework of quadratic gravity. This gap in the literature provides the primary impetus for the current study, which aims to address this overlooked aspect.   
 
In this study, we comprehensively analyze logarithmic corrections to the entropy of topological black holes within the framework of quadratic gravity. Our primary focus is to explore the influence of thermal fluctuations on the thermodynamic behaviour of these black holes. Beginning with deriving entropy corrections arising from thermal fluctuations, we establish the modified entropy expression incorporating logarithmic terms. A comparative graphical analysis is presented to elucidate the physical implications of these corrections, illustrating the variation of both corrected and uncorrected entropy with respect to the event horizon radius.
Furthermore, we extend our investigation to the system's internal energy by employing the first law of thermodynamics. The resulting corrections to internal energy, induced by thermal fluctuations, are computed and systematically analysed. These findings are substantiated through detailed graphical representations, offering critical insights into the modified thermodynamic structure of topological black holes under the influence of quantum statistical fluctuations. 

We proceed by computing the leading-order corrections to the Helmholtz free energy, which are crucial for understanding the thermodynamic behaviour of the black holes under perturbative thermal effects. To facilitate a comprehensive analysis, we plot graphs comparing the corrected Helmholtz free energy with its non-corrected counterpart as a function of the event horizon radius. These plots allow us to visually assess the impact of thermal fluctuations on the Helmholtz free energy.
Furthermore, we derive the corrected expression for the pressure of topological black holes, incorporating the effects of thermal fluctuations. Alongside this, we calculate the system's enthalpy, observing that it increases as a function of the horizon radius. This finding provides insights into how thermal fluctuations affect the thermodynamic properties of black holes.
We also investigate the impact of thermal fluctuations on the Gibbs free energy, examining how these fluctuations modify the Gibbs free energy compared to the uncorrected case. 
Finally, to assess the thermodynamic stability of the black holes, we perform a detailed analysis of the specific heat, incorporating corrections due to thermal fluctuations. This approach entails evaluating how such fluctuations modify the specific heat and, consequently, affect the stability criteria of the black holes. The resulting analysis reveals the significant role of thermal fluctuations in shaping the stability and overall thermodynamic behavior of topological black holes.

The paper is organised as follows: Section \ref{sec2} overviews topological black holes in $R^2$ gravity. Section \ref{sec3} focuses on the derivation and impact of logarithmic corrections to entropy due to thermal fluctuations. Section \ref{sec4} derives various corrected thermal quantities for topological black holes in $R^2$ gravity. Section \ref{sec5} discusses the stability of the black holes under thermal fluctuations. The final section presents conclusions and outlines directions for future research.

  \section{Topological black holes in  $ R^{2}$   gravity}\label{sec2}
Let us start by writing the  Lagrangian describing the pure $ R^{2}$   gravity as \cite{bg}
\begin{equation}
L_{0} =    \sqrt{•g} R^2.
\end{equation}
This Lagrangian results in the following field equation  \cite{bg}: 
\begin{equation}
 2RR_{\mu\nu} - \frac{1}{2•}R^{2}g_{\mu\nu} - 2\nabla_{\mu}\nabla_{\nu}R + 2g_{\mu\nu}\square R =0.
\end{equation}

The conformal transformation to the Einstein frame reads $ g_{\mu\nu} \rightarrow \Omega^{2}g_{\mu\nu}$.  Hence, all the solutions with $R = \Omega^{2} =0$ are excluded from the conformal mapping. 
The spherically symmetric black hole solutions are described by the following metric  \cite{bg}:
\begin{equation}
ds^{2} = - e^{2N(r)}dt^{2} + e^{-2N(r)}dr^{2} + r^{2}d\Sigma^{2}_k,\label{line}
\end{equation}
where 
\begin{equation}
N\left(r\right) = \frac{1}{2}\ln\left(k+\frac{a}{r} +br^{2}\right).
\end{equation}
Here, $a$ and $b$ are arbitrary constants. 
The configuration of spacetime is influenced by the parameters $ k, $ $ a, $ and $ b $. In particular, the zeros of  $ -g_{tt} = e^{2N}$ determine the location of the horizons. For $b=-\frac{\Lambda}{3} $, the above metric corresponds to asymptotic AdS black holes as
\begin{equation}
  -g_{tt} = k-\frac{\zeta M}{r}- \frac{\Lambda r^{2}}{3},\label{mm}
  \end{equation}
where parameter $\zeta$ depends on the Newton constant and   the
horizon space volume per unit radius. 
With this metric (\ref{mm}), the line element (\ref{line}) becomes
\begin{equation}
d s^{2}= -\left( k-\frac{•\zeta M}{r•}- \frac{\Lambda r^{2}}{3•}\right)dt^{2}+\frac{dr^{2}}{\left(k- \frac{\zeta M}{r}-\frac{\Lambda r^{2}}{•3}\right)} +r^{2}d\Sigma^{2}_{k},
\end{equation}
where $ d\Sigma^{2}_{k}=\frac{d\rho^{2}}{1-k \rho^{2}} + \rho^{2}d\phi^{2}$ and $k$ parametrizes the geometry of the horizon, e.g.,   $k =1  $ spherical,  $k= 0 $ flat or toroidal and  $ k= -1 $  
hyperbolic.

Once the metric function is determined, one can  calculate the temperature of the horizon  as \cite{bg}
  \begin{equation}
  T_{H} =   \frac{3r^{2}_{+}+ k l^{2}}{4 \pi  l^{2} r_{+}},
  \end{equation}
  where $\Lambda =-3/l^2$ is utilized. 
 
In general relativity (GR), the cosmological constant is treated as a fixed parameter in the action, leaving the black hole mass $M$ as the only variable that can, in principle, be adjusted. Accordingly, the internal energy depends solely on $M$ \cite{31}.  
In contrast, in $R^2$ gravity, the radius of the  AdS space is not fixed. Consequently, thermodynamic properties such as entropy and internal energy depend on $M$ and the AdS radius $l$.
The thermodynamic characteristics of topological black holes in  GR can be described using the Euclidean action. When  Euclidean action
 is well-defined, finite, and positive-definite, deriving the tree-level partition function becomes possible. This, in turn, allows for a formal definition of key thermodynamic quantities such as internal energy and entropy, analogous to their treatment within the canonical ensemble framework.
We calculate the original  entropy for the asymptotically AdS black hole \cite{bg}
  \begin{equation}
  S_{0} = \frac{96 \pi r_{+}^{2}}{l^{2}}.\label{eqn:S0}
  \end{equation}
  It is worth noting that this expression of entropy calculated from the  Euclidean procedure aligns with Wald's entropy \cite{xx}.
  \section{Logarithmic corrected entropy}\label{sec3}
Here, we briefly explain the origin of corrected entropy due to small statistical fluctuations as discussed in Ref. \cite{II5}. 
For this purpose,  we begin by specifying the general partition function for the given black hole as follows:
\begin{equation}
Z(\eta)=\int_{0}^{\infty}\rho(E)e^{-\eta E}dE,\label{par}
\end{equation}
where $\eta= 1/{T_{H}}$. The unit is set such that the Boltzmann constant takes the unit value. By using 
the inverse Laplace transformation, the above expression (\ref{par}) gives the density of state $\rho(E)$ as   
\begin{equation}
\label{eqn:II3}
\rho(E) =\frac{1}{2\pi i}\int_{\eta_{0}-i\infty}^{\eta_{0}+i\infty}e^{\mathcal{S}
(\eta)}
d\eta.
\end{equation}
Here, $\mathcal{S}(\eta)=\ln Z(\eta)+\eta E$ describes the exact black hole entropy. 
The evaluation of this complex integral (\ref{eqn:II3}) can be carried out using the steepest descent method, focusing on the saddle point if the  black holes are small enough  
such that $\left(\frac{\partial \mathcal{S}(\eta)}{\partial \eta}\right)_{\eta_{0}}=0$ and $
\left(\frac{\partial^{2} \mathcal{S}}{\partial \eta^{2}}\right)_{\eta_{0}}>0$. 

The expansion of  $\mathcal{S}(\eta)$ around the 
equilibrium point $\eta=\eta_{0}$, we have
\begin{equation}
\label{eqn:II4}
\mathcal{S}(\eta)=S_{0}+\frac{1}{2}(\eta - \eta_{0})^{2}\left(\frac{\partial^{2} \mathcal{S}}{\partial \eta^{2}}\right)_{\eta_{0}}+ \cdots,  
\end{equation}
where $S_{0}:=\mathcal{S}(\eta_0)$ and ${S}_0''=\left(\frac{\partial^{2} \mathcal{S}}{\partial \eta^{2}}\right)_{\eta_{0}}$.
In this context, $S_{0}:=\mathcal{S}(\eta_0)$ represents the equilibrium entropy of the black hole, as defined in expression (\ref{eqn:S0}). By carefully utilising the relationships outlined in expressions (\ref{eqn:II4}) and (\ref{eqn:II3}), one can derive the corresponding density of states. This density of states is a crucial quantity, as it characterises the distribution of microstates associated with the black hole at equilibrium. The derivation reinforces the connection between the black hole's macroscopic thermodynamic properties and its underlying quantum structure. It provides deeper insights into the statistical mechanics governing black hole entropy.
\begin{equation}
\rho(E)= \frac{e^{S_{0}}}{2 \pi i} \int_{\eta_{0}-i\infty}^{\eta_{0}+i\infty} e^{\frac{1}{2}(\eta - \eta_{0})^{2}{S}_0''} d\eta.
\end{equation}
Further simplification leads to
\begin{equation}
\rho(E) = \frac{e^{S_{0}}}{\sqrt{2 \pi {S}_0''}}.
\end{equation}
 For a system where no work is performed, the entropy correction resulting from thermal fluctuations can be approximated as \cite{II5}
\begin{eqnarray}
S&=&\ln \rho (E)\nonumber\\
&\approx & S_{0}-\alpha \ln   S_{0}.\label{eqn:II5}
\end{eqnarray}
Here, we do not fix the value of parameter $\alpha$  strictly  by the model as the value of $alpha$ may change when it accounts even higher order corrections.   Without the loss of generality,  we consider the value of $\alpha=1/2$ only to account the effects of fluctuation around the equilibrium on the various thermal quantities. The motivation of this choice is to only observe the change in nature of various thermal quantities under the effect of thermal fluctuations.

\begin{figure}[ht]
\centering{\includegraphics[scale=1]{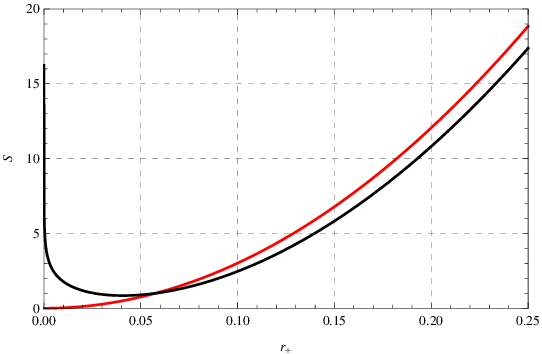}}
\caption{ Entropy versus Horizon radius. The red line denotes $ \alpha =0 $, and the black line denotes $ \alpha=0.5$. Here, $l=1$.}\label{fig1}
\end{figure}

  For the given value of equilibrium entropy (\ref{eqn:S0}), the corrected entropy for the topological black hole in $R^2$ gravity becomes
  \begin{equation}
  S =  \frac{96 \pi r_{+}^{2}}{l^{2}} - \alpha \ln  \frac{96 \pi r_{+}^{2}}{•l^{2}}.
  \end{equation}
  To study the effects of correction on the behaviour of entropy, we plot a graph as depicted in  Fig. \ref{fig1}. 
The figure demonstrates that the system's entropy remains positive across the examined range. A critical horizon radius can be identified, serving as a threshold that distinguishes different behaviours of the system. Specifically, for horizon radii greater than this critical value, the corrected entropy, much like the equilibrium entropy, increases as the horizon radius increases. This indicates that, for large topological black holes, the addition of thermal corrections continues to enhance the entropy as the black hole grows.

However, a notable deviation occurs when the horizon radius is smaller than the critical value. In this regime, the behaviour of the corrected entropy diverges from that of the equilibrium entropy. Unlike equilibrium entropy, which consistently increases with the horizon radius, the corrected entropy decreases as the horizon radius approaches the critical point from below. This suggests that thermal fluctuations destabilise smaller horizon radii, reducing the corrected entropy, contrary to the trend observed in the uncorrected case. This nuanced behaviour underscores the complex interplay between thermal corrections and black hole thermodynamics, particularly near the critical horizon radius.
  \section{Revised thermodynamic equations of state}\label{sec4}
\begin{figure}[h!]
\centering{\includegraphics[scale=1]{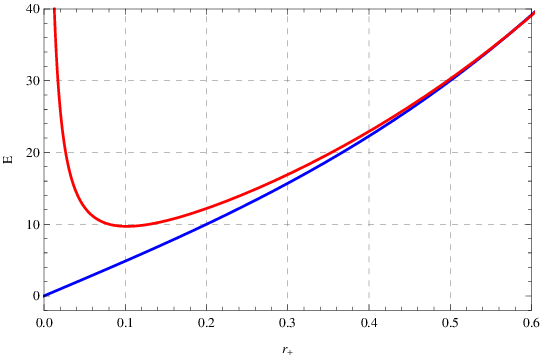}}
\caption{Internal energy versus horizon radius. Here, blue line denotes $\alpha=0 $,  and red line denotes $ \alpha=0.5$. Here, we set $l=1$, $k=1$.}\label{fig2}
  \end{figure}

 In this analysis, we derive the thermodynamic equations of state for a topological black hole within the framework of $R^2$ gravity, incorporating the effects of small thermal fluctuations into the system. With the established expressions for entropy and temperature, we can systematically calculate a range of other key thermodynamic quantities. For instance, the corrected internal energy, denoted by $E$, can be determined using the following fundamental relation: 
  \begin{eqnarray}
  E &=& \int T_{H}d S,\nonumber\\
&=&\frac{48}{l^{2}}\left(\frac{r_{+}^{3}}{l^{2}} +kr_{+}\right) -  {\alpha} \left(\frac{3r_{+}}{2\pi l^{2}} - \frac{k}{2\pi r_{+}}\right).
  \end{eqnarray}
  Now, to compare the corrected and uncorrected internal energy,   we plot  Fig.  \ref{fig2}.
  The figure illustrates the behaviour of both the corrected and equilibrium internal energy as functions of the horizon radius. It is evident that the equilibrium internal energy consistently increases with the radius of the event horizon. In contrast, the correction term introduced by thermal fluctuations significantly alters the behaviour of the internal energy for smaller black holes. Specifically, the corrected internal energy diverges for point-sized black holes, approaching an asymptotically considerable value. This deviation highlights the pronounced impact of thermal corrections on the internal energy in the regime of miniature black hole sizes.
  
\begin{figure}
  \centering{\includegraphics[scale=1]{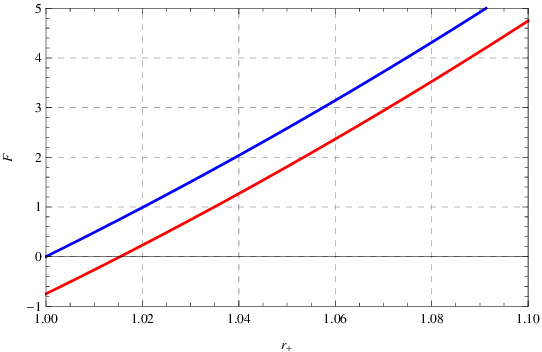}}
  \caption{The Helmholtz free energy versus horizon radius.     The blue line denotes $ \alpha=0 $, and the red line denotes $ \alpha=0.5 $. Here, we set $l = 1$, $k = 1$.}\label{fig3}
  \end{figure}

  The corrected Helmholtz free energy of a topological black hole in $R^2$ gravity  can be calculated with the help of the following standard relation: 
  \begin{equation}
  F = \int S dT_{H}.
  \end{equation}
 This leads to
  \begin{equation}
  F = \frac{24r_{+}^{3}}{l^{4}} - \frac{24kr_{+}}{l^{2}} - \frac{\alpha }{4\pi }  \left[\frac{3 r_{+}}{l^2}\left(\ln \frac{96 \pi r_{+}^{2}}{l^{2}} - 2\right)- \frac{ k}{ r_{+}}\left(\ln \frac{96 \pi r_{+}^{2}}{l^{2}} + 2\right)\right].
  \end{equation}
 In Fig. \ref{fig3}, we illustrate this expression to reveal the influence of thermal fluctuations on the Helmholtz free energy about the horizon radius. 
The graph reveals that the Helmholtz free energy consistently increases with the horizon radius. This trend holds regardless of thermal fluctuations, indicating that such fluctuations do not significantly alter the behaviour of the Helmholtz free energy for topological black holes within the context of $R^2$ gravity. The stability and monotonic increase of the Helmholtz free energy suggest that, even when thermal corrections are considered, the energy landscape of these black holes remains largely unaffected.

Furthermore, according to the area-entropy theorem, the entropy of a black hole is directly proportional to the area of its event horizon. This fundamental relationship allows us to extend our analysis and compute the topological black hole's corrected volume $ (V) $. The volume can be derived by considering how the entropy, influenced by thermal fluctuations, relates to the spatial extent of the black hole's horizon. The following expression provides the corrected volume of the topological black hole:  
 \begin{eqnarray}
 V &=& 4  \int S dr_+ =\frac{128 \pi r_{+}^{3}}{l^{2}} - 4 \alpha r_{+}\left(\ln\frac{96\pi r^{2}_{+}}{l^{2}} -2\right).
\end{eqnarray}  
Given that topological black holes are regarded as thermodynamic systems, it is essential to determine additional macroscopic parameters, such as pressure $(P)$. The pressure can be calculated using the standard definition provided below: 
\begin{eqnarray}
P &=&  -\frac{dF}{dV}=- \frac{1}{16\pi l^{2}r^{2}_{+}}\left( 3r_{+}^{2} - kl^{2}\right).
\end{eqnarray}
This expression represents the precise pressure of a topological black hole, considering the influence of minor statistical fluctuations. Notably, the absence of any term involving the parameter $\alpha$ indicates that the pressure remains independent of the correction parameter.
 \begin{figure}[ht]
\centering{\includegraphics[scale=1]{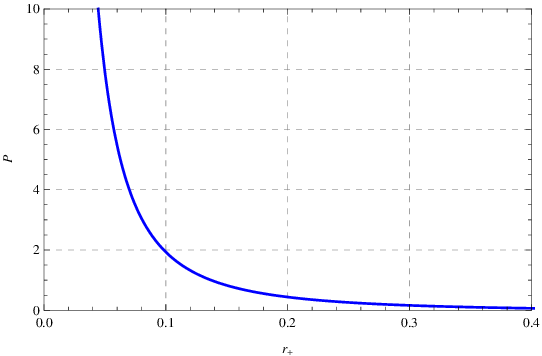}}
\caption{Pressure versus horizon radius. $k = 1 $and $l = 1$.}\label{fig4}
\end{figure}
The behaviour of pressure for this black hole is illustrated in Fig. \ref{fig4}, where it is evident that pressure diminishes as the size of the black hole increases.

Next, we focus on another crucial thermodynamic parameter: the system's enthalpy $ (H) $. Enthalpy, which represents the total heat content of the system, is defined as follows: 
\begin{eqnarray}
H &=& E + P V,\nonumber\\
& =& \frac{24  r_{+}^{3}}{l^4} +  \frac{56  r_{+}k}{l^2} -  \alpha  \left(\frac{ 3r_{+}^{2}-kl^2}{4\pi l^2 r_{+}^{2}}\ln\frac{96\pi r^{2}_{+}}{l^{2}} - \frac{ 3r_{+}^{2}-kl^2 }{2\pi l^2 r_{+}^{2}}\right).  
\end{eqnarray}
\begin{figure}
\centering{\includegraphics[scale=1]{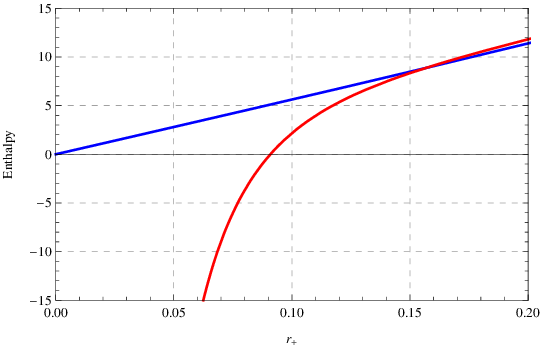}}
\caption{Enthalpy versus horizon radius. Here, blue line denotes $ \alpha=0 $ and red line denotes to $ \frac{1}{2} $. Here, we set  $k= 1$, $l =1$.}\label{fig5}
\end{figure}
To investigate the impact of thermal fluctuations on enthalpy, we present the graph in Fig. \ref{fig5}. The analysis reveals that, due to thermal corrections, the enthalpy of small black holes approaches a negative asymptotic value. This finding highlights the significant role that thermal fluctuations play in modifying the thermodynamic properties of these black holes, particularly at smaller scales.

\begin{figure}[h!]
 \centering{\includegraphics[scale=1]{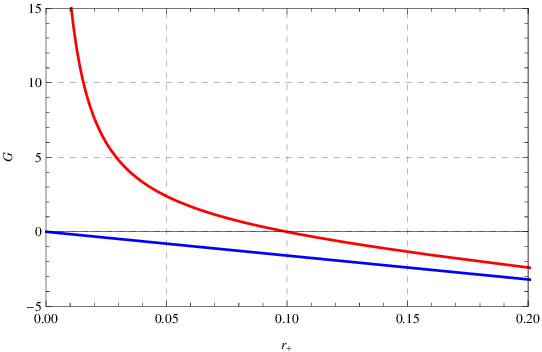}}
 \caption{Gibb's free energy versus horizon radius. Here, blue line denotes to $ \alpha = 0 $ and red line denotes to $  \frac{1}{2}$. Here, we set $ k = 1$, $l = 1$.}\label{fig6}
\end{figure}

The Gibbs free energy in thermodynamics represents the maximum amount of work that can be extracted from a closed system, such as black holes, under equilibrium conditions. To determine the corrected Gibbs free energy for topological black holes when thermal fluctuations are considered, we employ the following definition:
\begin{eqnarray}  
G & =&F + P V, \nonumber\\
& =& -\frac{16kr_{+}}{l^2} + \alpha \frac{k}{\pi r_{+}}.
\end{eqnarray}
To examine the behaviour of Gibbs free energy and the influence of thermal fluctuations, we present the graph illustrated in Fig. \ref{fig6}.
The plot reveals that the Gibbs free energy is negative for a topological black hole in thermal equilibrium. This negative value reflects the system's stability and thermodynamic properties at equilibrium. However, when considering the impact of thermal fluctuations, the Gibbs free energy of small black holes exhibits a significant shift. Specifically, it approaches large positive values asymptotically. This dramatic change underscores the substantial influence of thermal fluctuations on the Gibbs free energy, particularly for smaller black holes, where such fluctuations lead to a reversal from negative to positive values.
\section{Stability}\label{sec5}
\begin{figure}[h!]
\centering{
\includegraphics[scale=1]{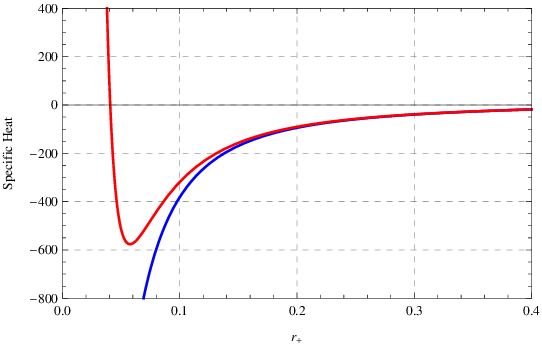}}
\caption{Specific heat versus horizon radius. Here, blue line denotes $ \alpha =0 $ and red line denotes to  $\alpha= \frac{1}{2} $. Here, we set $k = 1$, $l = 1$.}\label{fig7}
\end{figure}

To analyse the stability of a topological black hole within $R^2$ gravity, we compute the corrected specific heat by accounting for small thermal fluctuations. The behaviour of the specific heat is crucial, as its sign indicates the stability or instability of the black hole. To determine the corrected specific heat, we utilise the following relation: 
 \begin{equation}
   C_{v} = \frac{dE}{dT}.
\end{equation}
For the  topological black holes in $R^2$ gravity, this leads to
\begin{equation}
C_{v} = \frac{12}{\pi l^2}\left(\frac{9 r^2}{l^4} - \frac{k^2}{r^2}\right) -\frac{\alpha}{8 \pi ^2}\left(\frac{9}{l^4} - \frac{k^2}{r^4}\right).
\end{equation} 
To investigate how thermal fluctuations affect the signature of specific heat, we present the corresponding plot in Fig. \ref{fig7}.
 This plot reveals a significant shift in the stability characteristics of the black hole as a function of the horizon radius. Specifically, when thermal fluctuations are considered, the black hole undergoes two transitions in its state of stability: initially shifting from stable to unstable and subsequently from unstable back to stable as the horizon radius increases. This behaviour contrasts sharply with the equilibrium case, where the black hole transitions from unstable to stable only once. The introduction of thermal fluctuations thus introduces a more complex stability landscape, demonstrating their profound impact on the thermodynamic properties of the black hole.

 \section{Conclusion}\label{sec6}
 We have thoroughly examined the impact of thermal fluctuations on the topological black hole solutions in $R^2$ gravity, as detailed in references \cite{ab, cd, qw, er}. Our analysis reveals that thermal fluctuations induce a logarithmic correction to the entropy of these black holes. Notably, while this correction has a negligible effect on the entropy of large black holes, it significantly influences the entropy of smaller black holes.

Further, we have derived the corrected internal energy of the topological black hole using the first law of thermodynamics. Our findings indicate that, due to thermal fluctuations, the internal energy of small black holes becomes asymptotically large. Additionally, we have computed the corrected Helmholtz free energy, observing that it increases with the horizon radius. However, it remains consistently lower than its equilibrium counterpart. Significantly, thermal fluctuations do not alter the general behaviour of the Helmholtz free energy.

In our exploration of pressure and volume, we found that pressure becomes asymptotically large for small black holes, and interestingly, the pressure expression does not include a correction term. The corrected enthalpy, derived from our analysis, also increases the function of the horizon radius. However, thermal fluctuations cause the enthalpy to assume negative values for small black holes. We also evaluated the Gibbs free energy, reaching an asymptotically considerable value for small horizon radii. Meanwhile, the equilibrium Gibbs free energy tends towards zero. The Gibbs free energy decreases with increasing horizon radius.

Finally, to assess the stability of the black hole, we derived the corrected expression for specific heat. We discovered that, in equilibrium, the specific heat increases with the event horizon radius. Conversely, with thermal fluctuations, the black hole undergoes two transitions in stability: first from stable to unstable and then from unstable back to stable as the horizon radius increases. In the equilibrium scenario, the black hole transitions from unstable to stable only once. Future research will explore the effects of non-perturbative corrections on the thermodynamics of topological black holes, which promises to offer further insights into their complex behaviour.  
\section*{Acknowledgement}
This research was funded by the Science Committee of the Ministry of Science and Higher Education of the
Republic of Kazakhstan (Grant No. AP22682760). D.V.S. would like to thank DST-SERB for the project no. EEQ/2022/00824.

%======================================================
\end{document}